\documentclass[11pt]{article}
\begin{document}

\title{On the evaluation of the evolution operator $Z_{\rm
Reg}(R_2,R_1)$ in the Diakonov--Petrov approach to the Wilson
loop}

\author{M. Faber\thanks{E--mail: faber@kph.tuwien.ac.at, Tel.:
+43--1--58801--14261, Fax: +43--1--58801--14299} ,
A. N. Ivanov\thanks{E--mail: ivanov@kph.tuwien.ac.at, Tel.:
+43--1--58801--14261, Fax: +43--1--58801--14299}~$^{\ddagger}$ , N. I.
Troitskaya\thanks{Permanent Address: State
Technical University, Department of Nuclear Physics, 195251
St. Petersburg, Russian Federation}}

\date{\today}

\maketitle

\begin{center}
{\it Institut f\"ur Kernphysik, Technische Universit\"at Wien, \\ 
Wiedner Hauptstr. 8--10, A--1040 Vienna, Austria}
\end{center}

\begin{abstract}
We evaluate the evolution operator $Z_{\rm Reg}(R_2,R_1)$ introduced
by Diakonov and Petrov for the definition of the Wilson loop in terms
of a path integral over gauge degrees of freedom. We use the procedure
suggested by Diakonov and Petrov (Phys. Lett. B224 (1989) 131) and
show that the evolution operator vanishes.
\end{abstract}

\begin{center}
\end{center}

\section*{Path integral for the evolution operator $Z_{\rm Reg}(R_2,R_1)$ } 
\setcounter{equation}{0}

\hspace{0.2in} In Ref.[1] for the representation of the Wilson loop in
terms of the path integral over gauge degrees of freedom Diakonov and
Petrov used the functional $Z(R_2,R_1)$ defined by (see Eq.(8) of
Ref.[1])
\begin{eqnarray}\label{label1}
Z(R_2,R_1) =
\int\limits^{R_2}_{R_1}DR(t)\,\exp\Bigg(iT\int\limits^{t_2}_{t_1}\,{\rm
Tr}\,(iR\,\dot{R}\,\tau_3)\Bigg),
\end{eqnarray}
where $\dot{R} = dR/dt$ and $T = 1/2,1,3/2,\ldots$ is the colour
isospin quantum number. According to Diakonov and Petrov $Z(R_2,R_1)$
should be regularized by the analogy to an axial--symmetric top. The
regularized expression of $Z(R_2,R_1)$ has been determined in Eq.(9)
of Ref.[1] and reads
\begin{eqnarray}\label{label2}
Z_{\rm Reg}(R_2,R_1) =
\int\limits^{R_2}_{R_1}DR(t)\,\exp\Bigg(i\int\limits^{t_2}_{t_1}
\Big[\frac{1}{2}\,I_{\perp}\,(\Omega^2_1 + \Omega^2_2) +
\frac{1}{2}\,I_{\parallel}\,\Omega^2_3 + T\,\Omega_3\Big]\Bigg),
\end{eqnarray}
where $\Omega_a = i\,{\rm Tr}(R\,\dot{R}\,\tau_a)$ are angular
velocities of the top, $\tau_a$ are Pauli matrices $a=1,2,3$,
$I_{\perp}$ and $I_{\parallel}$ are the moments of inertia of the top
which should be taken to zero. According to the prescription of
Ref.[1] one should take first the limit $I_{\parallel} \to 0$ and then
$I_{\perp} \to 0$. For the confirmation of the result, given in
Eq.(13) of Ref.[1],
\begin{eqnarray}\label{label3}
Z_{\rm Reg}(R_2,R_1) = (2T + 1)\,D^T_{TT}(R_2R^{\dagger}_1) = (2T +
1)\,D^T_{-T-T}(R_1R^{\dagger}_2),
\end{eqnarray}
where $D^T(U)$ is a Wigner rotational matrix in the representation
$T$, Diakonov and Petrov suggested to evaluate the evolution operator
(\ref{label2}) explicitly via the discretization of the path integral
over $R$.  The discretized form of the path integral Eq.(\ref{label2})
is given by Eq.(14) of Ref.[1] and reads
\begin{eqnarray}\label{label4}
\hspace{-0.5in}&&Z_{\rm Reg}(R_{N+1},R_0) = \lim_{\begin{array}{c} N\to
\infty\\ \delta \to 0\end{array}}{\cal N}\int\prod^{N}_{n=1}dR_n\nonumber\\ 
\hspace{-0.5in}&&\times\,\exp\Bigg[\sum^{N}_{n=0}
\Bigg(-\,i\,\frac{I_{\perp}}{2\delta}\,\Big[({\rm Tr}\,V_n\tau_1)^2 +
({\rm Tr}\,V_n\tau_2)^2\Big] - \,i\,\frac{I_{\parallel}}{2\delta}\,({\rm
Tr}\,V_n\tau_3)^2 -\,T\,({\rm Tr}\,V_n\tau_3)\Bigg)\Bigg],
\end{eqnarray}
where $R_n = R(s_n)$ with $s_n = t_1 + n\,\delta$ and $V_n = R_n
R^{\dagger}_{n+1}$ are the relative orientations of the top at
neighbouring points [1]. The normalization factor ${\cal N}$ is
determined by
\begin{eqnarray}\label{label5}
{\cal N} = \Bigg(\frac{I_{\perp}}{2\pi i
\delta}\sqrt{\frac{I_{\parallel}}{2\pi i \delta}}\,\Bigg)^{N+1}.
\end{eqnarray}
(see Eq.(19) of Ref.[1]). Following the prescription of Ref.[1] one
should take the limits $\delta \to 0$ and $I_{\parallel}, I_{\perp}
\to 0$ but keeping the ratios $I_i/\delta$, where ($ i= {\parallel},
{\perp}$), much greater than unity, $I_i/\delta \gg 1$.

The main point of the evaluation of the path integral is to show that
the evolution operator $Z_{\rm Reg}(R_2,R_1)$ given by the path
integral (\ref{label2}) reduces to the representation in the form of
{\it a sum over possible intermediate states}, i.e. eigenfunctions of
the axial--symmetric top [1]
\begin{eqnarray}\label{label6}
Z_{\rm Reg}(R_2,R_1) =\sum^{\infty}_{J=0}\sum^J_{m = - J}(2J + 1)\,
D^J_{m m}(R_2R^{\dagger}_1)\,e^{\textstyle -i(t_2-t_1)\,E_{J m}},
\end{eqnarray}
(see Eq.(12) of Ref.[1]), where $E_{J m}$ are the eigenvalues of the
Hamiltonian of the axial--symmetric top 
\begin{eqnarray}\label{label7}
E_{J m} = \frac{J(J+1) - m^2}{2 I_{\perp}} + \frac{(m - T)^2}{2
I_{\parallel}}
\end{eqnarray}
(see Eq.(11) of Ref.[1]).

According to Diakonov$^{\prime}$s and Petrov$^{\prime}$s statement the
integral has a saddle--point at $V_n \simeq 1$. For the calculation of
the integral around the saddle--point Diakonov and Petrov suggested
the following procedure. Let us denote the exponent of
Eq.(\ref{label4}) as
\begin{eqnarray}\label{label8}
f[V_n] = -\,i\,\frac{I_{\perp}}{2\delta}\,\Big[({\rm Tr}\,V_n\tau_1)^2 +
({\rm Tr}\,V_n\tau_2)^2\Big] - \,i\,\frac{I_{\parallel}}{2\delta}\,({\rm
Tr}\,V_n\tau_3)^2 -\,T\,({\rm Tr}\,V_n\tau_3) 
\end{eqnarray}
and represent the exponential in the following form
\begin{eqnarray}\label{label9}
e^{\textstyle f[V_n]} =
\sum^{\infty}_{J=0}\sum^{J}_{p=-J}\sum^{J}_{q=-J}(2J +
1)\lambda^J_{pq}D^J_{pq}(V_n).
\end{eqnarray}
The coefficients $\lambda^J_{pq}$ are given by
\begin{eqnarray}\label{label10}
\lambda^J_{pq} = \int dU_n\,D^J_{qp}(U^{\dagger}_n)\,e^{\textstyle f[U_n]}.
\end{eqnarray}
Substituting Eq.(\ref{label10}) in Eq.(\ref{label9}) we get the identity
\begin{eqnarray}\label{label11}
e^{\textstyle f[V_n]} =
\sum^{\infty}_{J=0}\sum^{J}_{p=-J}\sum^{J}_{q=-J}(2J +
1)\,D^J_{pq}(V_n)\, \int dU_n\,D^J_{qp}(U^{\dagger}_n)\,e^{\textstyle
f[U_n]}.
\end{eqnarray}
Let us show that Eq.(\ref{label11}) is the identity. For this aim we
have to use the relation
\begin{eqnarray}\label{label12}
\sum^{\infty}_{J=0}\sum^{J}_{p=-J}\sum^{J}_{q=-J}(2J + 1)\,
D^J_{pq}(V_n)\,D^J_{qp}(U^{\dagger}_n) =
\sum^{\infty}_{J=0}(2J+1)\,\chi_J[V_nU^{\dagger}_n].
\end{eqnarray}
By using Eq.(\ref{label12}) the r.h.s. of Eq.(\ref{label11}) reads
\begin{eqnarray}\label{label13}
\int dU_n\,e^{\textstyle
f[U_n]}\,\sum^{\infty}_{J=0}(2J+1)\,\chi_J[V_nU^{\dagger}_n] = \int
dU_n\,e^{\textstyle f[U_n]}\,\delta(V_nU^{\dagger}_n) = e^{\textstyle
f[V_n]},
\end{eqnarray}
where $\delta(V_nU^{\dagger}_n)$ is a $\delta$--function defined by 
\begin{eqnarray}\label{label14}
\sum^{\infty}_{J=0}(2J+1)\,\chi_J[V_nU^{\dagger}_n] =
\delta(V_nU^{\dagger}_n).
\end{eqnarray}
The important consequence of these steps is that $dU_n$ as well as
$dV_n$ is a standard Haar measure normalized to unity
\begin{eqnarray}\label{label15}
\int dU_n = \int dV_n = 1.
\end{eqnarray}
This point alters crucially the results of Ref.[1].

Inserting the expansion Eq.(\ref{label11}) in the r.h.s. of
Eq.(\ref{label4}) we obtain
\begin{eqnarray}\label{label16}
\hspace{-0.5in}&&Z_{\rm Reg}(R_{N+1},R_0) = \lim_{\begin{array}{c}
N\to \infty\\ \delta \to 0\end{array}}{\cal N}\int\ldots\int
dR_1\,dR_2\ldots dR_{N-1}\,dR_N\nonumber\\
\hspace{-0.5in}&&\times\,\sum^{\infty}_{J_0=0}
\sum^{J_0}_{p_0=-J_0}\sum^{J_0}_{q_0=-J_0}(2J_0
+ 1)\,D^{J_0}_{p_0q_0}(R_0R^{\dagger}_1)\, \int
dU_0\,D^{J_0}_{q_0p_0}(U^{\dagger}_0)\,e^{\textstyle
f[U_0]}\nonumber\\
\hspace{-0.5in}&&\times\,\sum^{\infty}_{J_1=0}
\sum^{J_1}_{p_1=-J_1}\sum^{J_1}_{q_1=-J_1}(2J_1
+ 1)\,D^{J_1}_{p_1q_1}(R_1R^{\dagger}_2)\, \int
dU_1\,D^{J_1}_{q_1p_1}(U^{\dagger}_1)\,e^{\textstyle
f[U_1]}\nonumber\\
\hspace{-0.5in}&&\times\,\sum^{\infty}_{J_2=0}
\sum^{J_2}_{p_2=-J_2}\sum^{J_2}_{q_2=-J_2}(2J_2 + 1)
\,D^{J_2}_{p_2q_2}(R_2R^{\dagger}_3)\, \int
dU_2\,D^{J_2}_{q_2p_2}(U^{\dagger}_2)\,e^{\textstyle
f[U_2]}\nonumber\\ \hspace{-0.5in}&&\times\,\ldots\nonumber\\
\hspace{-0.5in}&&\times\,\sum^{\infty}_{J_N=0} \sum^{J_N}_{p_N
=-J_N}\sum^{J_N}_{q_N =-J_N}(2J_N +
1)\,D^{J_N}_{p_Nq_N}(R_NR^{\dagger}_{N+1})\, \int
dU_N\,D^{J_N}_{q_Np_N}(U^{\dagger}_N)\,e^{\textstyle f[U_N]}
\end{eqnarray}
Integrating over $R_n\,(n=1,2,\ldots,N)$ and using the
orthogonality relation for the group elements we arrive at the
expression
\begin{eqnarray}\label{label17}
\hspace{-0.5in}&&Z_{\rm Reg}(R_{N+1},R_0) = \lim_{\begin{array}{c}
N\to \infty\\ \delta \to 0\end{array}}
\sum^{\infty}_{J=0}\sum^{J}_{p=-J}\sum^{J}_{q=-J}(2J +
1)\,D^J_{pq}(R_0R^{\dagger}_{N+1})\,{\cal N}\int
dU_0\,D^J_{qp}(U^{\dagger}_0)\,e^{\textstyle
f[U_0]}\nonumber\\
\hspace{-0.5in}&&\times \int
dU_1\,D^J_{qp}(U^{\dagger}_1)\,e^{\textstyle f[U_1]}\int
dU_2\,D^J_{qp}(U^{\dagger}_2)\,e^{\textstyle f[U_2]}\ldots\int
dU_N\,D^J_{qp}(U^{\dagger}_N)\,e^{\textstyle f[U_N]}=\nonumber\\
\hspace{-0.5in}&&=\lim_{\begin{array}{c} N\to \infty\\ \delta \to
0\end{array}}\sum^{\infty}_{J=0}\sum^{J}_{p=-J}
\sum^{J}_{q=-J}(2J + 1)\,D^J_{pq}(R_0R^{\dagger}_{N+1})\,[Z^J_{qp}]^{N+1},
\end{eqnarray}
where $Z^J_{qp}$ is defined by
\begin{eqnarray}\label{label18}
Z^J_{qp} = \frac{I_{\perp}}{2\pi i
\delta}\sqrt{\frac{I_{\parallel}}{2\pi i \delta}}\int
dU\,D^J_{qp}(U^{\dagger})\,e^{\textstyle f[U]}.
\end{eqnarray}
Recall that $dU$ is the Haar measure normalized to unity
Eq.(\ref{label15}).

For the subsequent evaluation of the integral over $U$ we follow
Diakonov and Petrov and use
\begin{eqnarray}\label{label19}
U = e^{\textstyle i\,\frac{1}{2}\,\vec{\omega}\cdot \vec{\tau}}
\end{eqnarray}
for the fundamental representation and
\begin{eqnarray}\label{label20}
D^J_{qp}(U^{\dagger}) = \Big(e^{\textstyle -\,i\,\vec{\omega}\cdot
\vec{T}}\,\Big)_{qp}
\end{eqnarray}
for $J\not= 1/2$. In the parameterization (\ref{label19}) the Haar
measure $dU$ reads
\begin{eqnarray}\label{label21}
dU =
\frac{d\omega_1d\omega_2d\omega_3}{16\pi^2}\,\left(\frac{2}{\omega}\,
\sin\frac{\omega}{2}\right)^2,
\end{eqnarray}
where $\omega = \sqrt{\omega^2_1 + \omega^2_2 + \omega^2_3}$.
According to the Diakonov and Petrov point of view the integral over
$U$ calculated in the limit $I_{\parallel}/\delta, I_{\perp}/\delta
\to \infty$ has a saddle point at $U\simeq 1$\,\footnote{Below we do
not pay attention to the factor $1/16\pi^2$ that has to be included in
the normalization factor ${\cal N}$ in the form
$(16\pi^2)^{N+1}$.}. Expanding the integrand around the saddle--point,
keeping only quadric terms and {\bf neglecting the contribution of the
terms coming from the Haar measure}, we get
\begin{eqnarray}\label{label22}
Z^J_{qp} &=& \frac{I_{\perp}}{2\pi i
\delta}\sqrt{\frac{I_{\parallel}}{2\pi i
\delta}}\int\limits^{\infty}_{-\infty}d\omega_1
\int\limits^{\infty}_{-\infty}d\omega_2\int\limits^{\infty}_{-\infty}
d\omega_3\,\exp\, \Bigg\{i\frac{I_{\perp}}{2\delta}\,(\omega^2_1 +
\omega^2_2) +
i\frac{I_{\parallel}}{2\delta}\,\omega^2_3\Bigg\}\nonumber\\ &&\times
\Bigg[\delta_{qp} - \frac{1}{2}\,[\omega^2_1(T^2_1)_{qp} +
\omega^2_2(T^2_2)_{qp}] - \frac{1}{2}\,\omega^2_3\,((T_3+ 
T)^2)_{qp}\Bigg].
\end{eqnarray}
Integrating over $\omega_a\,(a=1,2,3)$ we arrive at the expression
\begin{eqnarray}\label{label23}
Z^J_{qp} &=& \delta_{qp} - i\,\delta\,\Bigg[\frac{(T^2_1 + T^2_2)_{qp}}{2I_{\perp}}
+ \frac{((T_3 + T)^2)_{qp}}{2I_{\parallel}}\Bigg]=\nonumber\\
&=&\delta_{qp}\,\Bigg\{1 -i\,\delta\,\Bigg[\frac{(J(J+1) - p^2)}{2I_{\perp}}
+ \frac{(p + T)^2}{2I_{\parallel}}\Bigg]\Bigg\}.
\end{eqnarray}
This agrees with the result obtained by Diakonov and Petrov (see
Eq.(18) of Ref.[1])

Substituting Eq.(\ref{label23}) in Eq.(\ref{label17}) we obtain the
evolution operator $Z_{\rm Reg}(R_0R^{\dagger}_{N+1})$ defined by
\begin{eqnarray}\label{label24}
\hspace{-0.5in}&&Z_{\rm Reg}(R_{N+1},R_0) = \lim_{\begin{array}{c}
N\to \infty\\ \delta \to
0\end{array}}\sum^{\infty}_{J=0}\sum^{J}_{p=-J}\sum^{J}_{q=-J}(2J +
1)\,D^J_{pq}(R_0R^{\dagger}_{N+1})\,[Z^J_{qp}]^{N+1}=\nonumber\\
\hspace{-0.5in}&&=\lim_{\begin{array}{c} N\to \infty\\ \delta \to
0\end{array}}\sum^{\infty}_{J=0}\sum^{J}_{p = -J}(2J +
1)\,D^J_{pp}(R_0R^{\dagger}_{N+1})\,\Bigg\{1
-i\,\delta\,\Bigg[\frac{(J(J+1) - p^2)}{2I_{\perp}} + \frac{(p +
T)^2}{2I_{\parallel}}\Bigg]\Bigg\}^{N+1}=\nonumber\\
\hspace{-0.5in}&&=\lim_{N\to \infty}\sum^{\infty}_{J=0}\sum^{J}_{p =
-J}(2J + 1)\,D^J_{pp}(R_0R^{\dagger}_{N+1})\,\Bigg\{1
-i\,\frac{t_2-t_1}{N+1}\,\Bigg[\frac{(J(J+1) - p^2)}{2I_{\perp}} +
\frac{(p + T)^2}{2I_{\parallel}}\Bigg]\Bigg\}^{N+1},
\end{eqnarray}
where we have used the definition of $\delta$: $\delta = (t_2 -
t_1)/(N+1)$ [1].

Taking the limit $N\to \infty$ we get
\begin{eqnarray}\label{label25}
\hspace{-0.1in}Z_{\rm Reg}(R_{\infty},R_0) &=&  \sum^{\infty}_{J=0}
\sum^{J}_{p= -J}(2J +
1)\,D^J_{pp}(R_0R^{\dagger}_{\infty})\nonumber\\
&&\times\,\exp\Bigg\{-i(t_2-t_1)\,\Bigg[\frac{(J(J+1)
- p^2)}{2I_{\perp}} + \frac{(p + T)^2}{2I_{\parallel}}\Bigg]\Bigg\}.
\end{eqnarray}
Replacing $R_0 \to R_1$ and $R^{\dagger}_{\infty} \to R^{\dagger}_2$
we arrive at the expression 
\begin{eqnarray}\label{label26}
Z_{\rm Reg}(R_2,R_1) &=& \sum^{\infty}_{J=0}\sum^{J}_{p= -J}(2J +
1)\,D^J_{pp}(R_1R^{\dagger}_2)\nonumber\\
&&\times\,\exp\Bigg\{-i\,(t_2-t_1)\,
\Bigg[\frac{(J(J+1) - p^2)}{2I_{\perp}} + \frac{(p +
T)^2}{2I_{\parallel}}\Bigg]\Bigg\}.
\end{eqnarray}
This expression coincides fully with the result obtained by Diakonov
and Petrov (see Eq.(22) of Ref.[1]) and reproduces the expansion of
the evolution operator (\ref{label6}) (see Eq.(12) of
Ref.[1]).

Now taking the limits $I_{\parallel} \to 0$ and $I_{\perp} \to 0$ we
have to keep the term $-p = J = T$ [1] and obtain
\begin{eqnarray}\label{label27}
Z_{\rm Reg} (R_2, R_1) =
(2T+1)\,D^T_{-T-T}(R_1R^{\dagger}_2)\,\exp\Bigg[ -i(t_2-t_1)\,\frac{T}{2
I_{\perp}}\Bigg].
\end{eqnarray}
In the limit $I_{\perp} \to 0$ due to this strongly oscillating factor
the r.h.s. of Eq.(\ref{label27}) vanishes. This point has been
discussed in detail in Refs.[2,3]. Such a vanishing of the evolution
operator confirms the statement in Refs.[2,3] that the path integral
representation of the Wilson loop by Diakonov and Petrov is
erroneous.

We would like to accentuate that following Diakonov$^{\prime}$s and
Petrov$^{\prime}$s evaluation of the integral over $U$ we have not
taken into account the contribution of the Haar measure. From the Haar
measure (\ref{label21}) we should get an additional contribution
\begin{eqnarray}\label{label28}
dU =
\frac{d\omega_1d\omega_2d\omega_3}{16\pi^2}\,
\left(\frac{2}{\omega}\,\sin\frac{\omega}{2}\right)^2
= \frac{d\omega_1d\omega_2d\omega_3}{16\pi^2}\,\Big( 1 -
\frac{1}{12}\,(\omega^2_1 + \omega^2_2 + \omega^2_3)\Big).
\end{eqnarray}
This changes the value $Z^J_{qp}$ in Eq.(\ref{label23}) as follows
\begin{eqnarray}\label{label29}
Z^J_{qp} = \delta_{qp}\,\Bigg\{1
-i\,\delta\,\frac{1}{12}\,\Bigg(\frac{2}{I_{\perp}}+
\frac{1}{I_{\parallel}}\Bigg) - i\,\delta\,\Bigg[\frac{(J(J+1) -
p^2)}{2I_{\perp}} + \frac{(p + T)^2}{2I_{\parallel}}\Bigg]\Bigg\}.
\end{eqnarray}
However, it is not the complete set of contributions of order
$O(\delta/I_{\perp})$ and $O(\delta/I_{\parallel})$ to $Z^J_{qp}$. In
order to take into account all of them we have to expand too the
exponential $\exp\,f[U]$ keeping the terms of order
$\omega^4_1I_{\perp}/\delta$, $\omega^4_2I_{\perp}/\delta$,
$\omega^4_3I_{\parallel}/\delta$ and so on.  The corresponding expansion
of the exponential $\exp\,f[U]$ reads
\begin{eqnarray}\label{label30}
\exp\,f[U] &=& \exp\, \Bigg\{i\frac{I_{\perp}}{2\delta}\,(\omega^2_1 +
\omega^2_2) +
i\frac{I_{\parallel}}{2\delta}\,\omega^2_3\Bigg\}\nonumber\\
&&\times\,\Bigg[1 - i\frac{I_{\perp}}{24\delta}(\omega^2_1 +
\omega^2_2)^2 - i\frac{I_{\perp} + I_{\parallel}}{24\delta}(\omega^2_1
+ \omega^2_2)\,\omega^2_3 - i\frac{I_{\parallel}}{24\delta}\omega^4_3
+\ldots\Bigg],
\end{eqnarray}
where ellipses denote the terms that have been taken into account in
(\ref{label22}).

The contribution of the terms in Eq.(\ref{label30}) changes $Z^J_{qp}$
(\ref{label29}) as follows
\begin{eqnarray}\label{label31}
Z^J_{qp} = \delta_{qp}\,\Bigg\{1
+ i\,\delta\,\frac{1}{8}\,\Bigg(\frac{2}{I_{\perp}}+
\frac{1}{I_{\parallel}}\Bigg) - i\,\delta\,\Bigg[\frac{(J(J+1) -
p^2)}{2I_{\perp}} + \frac{(p + T)^2}{2I_{\parallel}}\Bigg]\Bigg\}.
\end{eqnarray}
This describes the total contribution of the terms of order
$O(\delta/I_{\perp})$ and $O(\delta/I_{\parallel})$. Due to
Eq.(\ref{label31}) the evolution operator reads
\begin{eqnarray}\label{label32}
&&Z_{\rm Reg}(R_2,R_1) =
\exp\Bigg\{i\,(t_2-t_1)\,\frac{1}{8}\,\Bigg(\frac{2}{I_{\perp}}+
\frac{1}{I_{\parallel}}\Bigg)\Bigg\}\nonumber\\
&&\times\sum^{\infty}_{J}\sum^{J}_{p= -J}(2J +
1)\,D^J_{pp}(R_1R^{\dagger}_2)\,\exp\Bigg\{-i\,(t_2-t_1)\,
\Bigg[\frac{(J(J+1) - p^2)}{2I_{\perp}} + \frac{(p +
T)^2}{2I_{\parallel}}\Bigg]\Bigg\}.
\end{eqnarray}
{\bf Hence, the evaluation of the path integral (\ref{label2}) with
the correct account for all contributions of order
$O(\delta/I_{\perp})$ and $O(\delta/I_{\parallel})$ around the
saddle--point, including the contributions of the Haar measure and the
terms of order $\omega^4_1I_{\perp}/\delta$,
$\omega^4_2I_{\perp}/\delta$, $\omega^4_3I_{\parallel}/\delta$ and so on,
leads to a result that differs fully from the expansion (\ref{label6})
derived from the quantum mechanical consideration of $Z_{\rm Reg}(R_2,
R_1)$ in terms of eigenfunctions of the axial--symmetric top. This
means that the path integral (\ref{label2}) representing the evolution
operator $Z_{\rm Reg}(R_2, R_1)$ has no relation to the
axial--symmetric top and predicts a completely different energy
spectrum than that given by Eq.(\ref{label7}) for the quantum
axial--symmetric top. In the limit $I_{\parallel} \to 0$ and
$I_{\perp} \to 0$ the evolution operator vanishes by virtue of the
strongly oscillating factors.}

Thus, the only well defined magnitude of the evolution operator is
zero. This confirms fully the results obtained in Refs.[2,3] that the
evolution operator $Z_{\rm Reg}(R_2, R_1)$ vanishes and the path
integral representation of the Wilson loop suggested by Diakonov and
Petrov in terms of the evolution operator $Z_{\rm Reg}(R_2, R_1)$ is
erroneous. All of these statements are completely applicable to the
results discussed by Diakonov and Petrov in their recent manuscript
hep--lat/0008004 [4].

\section*{Acknowledgement}

\hspace{0.2in} This investigation has been initiated by Oleg Borisenko
having a great interest in the path integral representation of the
Wilson loop suggested by Diakonov and Petrov in Refs.[1,4].

\end{document}